\documentclass{svproc}
\usepackage{url}

\usepackage{graphicx}
\usepackage{multirow}

\begin{document}
\mainmatter              
\title{A Comparative Study of Machine Learning Models for Tabular Data Through Challenge of Monitoring Parkinson’s Disease Progression Using Voice Recordings}
\titlerunning{A Comparative Study of ML Models for Tabular Data}  
%
\author{Mohammadreza Iman\inst{1}\thanks{These authors contributed equally.} \and Amy Giuntini\inst{2}\footnotemark[1]\and Hamid Reza Arabnia\inst{1}\and
Khaled Rasheed\inst{2}}
\authorrunning{Mohammadreza Iman et al.} 
%
\tocauthor{Mohammadreza Iman, Amy Giuntini, Hamid Reza Arabnia, Khaled Rasheed}
\institute{Department of Computer Science, 
Franklin College of Arts and Sciences
University of Georgia
Athens, GA, USA,\\
\email{mohammadreza.iman@hotmail.com},\\ \email{hra@uga.edu}
\and
Institute for Artificial Intelligence, Franklin College of Arts and Sciences
University of Georgia
Athens, GA, USA,\\
\email{algiuntini@gmail.com},\\
\email{khaled@uga.edu}
}
\maketitle              

\begin{abstract}
People with Parkinson’s disease must be regularly monitored by their physician to observe how the disease is progressing and potentially adjust treatment plans to mitigate the symptoms. Monitoring the progression of the disease through a voice recording captured by the patient at their own home can make the process faster and less stressful. Using a dataset of voice recordings of 42 people with early-stage Parkinson’s disease over a time span of 6 months, we applied multiple machine learning techniques to find a correlation between the voice recording and the patient’s motor UPDRS score. We approached this problem using a multitude of both regression and classification techniques. Much of this paper is dedicated to mapping the voice data to motor UPDRS scores using regression techniques in order to obtain a more precise value for unknown instances. Through this comparative study of variant machine learning methods, we realized some old machine learning methods like trees outperform cutting edge deep learning models on numerous tabular datasets.
\keywords{Parkinson’s disease, Machine Learning, Tabular Data, Feature Selection, Regression, Ensemble Models, Multi-instance Learning}
\end{abstract}
\section{Introduction}
Parkinson's disease is a neurodegenerative disorder, affecting the neurons in the brain that produce dopamine. Parkinson's disease can cause a range of symptoms, particularly the progressive deterioration of motor function [1] [2] [3]. When diagnosed with Parkinson’s disease, a person’s health may deteriorate rapidly, or they may experience comparatively milder symptoms if the disease progresses more slowly. In our research, we are mainly concerned with how Parkinson's disease can affect speech characteristics. People with Parkinson's may display dysarthria or problems with articulation, and they may also be affected by dysphonia, an impaired ability to produce vocal sounds normally. Dysphonia may be exhibited by soft speech, breathy voice, or vocal tremor [1] [2] [3].

People with Parkinson's disease do not always experience noticeable symptoms at the earliest stages, and therefore, the disease is often diagnosed at a later stage. As there is not currently a cure, people diagnosed with Parkinson's must rely on treatments to alleviate symptoms, which is most effective with early treatment. Once a diagnosis is made, the patient must regularly visit their physician to monitor the disease and the effectiveness of treatment. Monitoring the progression of the disease through a voice recording captured by the patient at their own home can make the process faster and less stressful for the patient. The possibility of lessening the frequency of doctor visits can be cost-effective and allow the patient to follow a more flexible schedule [1] [2] [3].

One of the most prominent methods of quantifying the symptoms of Parkinson's disease is the Unified Parkinson's Disease Rating Scale (UPDRS), which was first developed in the 1980s. The scale consists of four parts: intellectual function and behavior, ability to carry out daily activities, motor function examination, and motor complications. Each part is composed of questions or tests where either the patient or clinician will give a score with 0 denoting normal function and the maximum number denoting severe impairment. The clinician calculates a UPDRS score for each section as well as a total UPDRS score that can range from 0 to 176. In this work, we are concerned with the motor UPDRS, which can range from 0 to 108.

All the machine learning techniques and data analysis in this project have been done using the Waikato Environment for Knowledge Analysis (Weka), free software developed at the University of Waikato [4]. Weka is a compilation of machine learning algorithms written in Java. All the applied methods in this study are based on 10-fold cross-validation.

\subsection{Dataset}

We obtained the data from UC Irvine’s machine learning repository. The data consists of a total of 5,875 voice recordings from 42 patients with early-stage Parkinson’s disease over the course of 6 months. Voice recordings were taken for each subject weekly, and a clinician determined the subject’s UPDRS scores at the onset of the trial, at three months, and at six months. The scores were then linearly interpolated for the remaining voice recordings [2].

The raw data from the original paper had 132 attributes, but the publicly available data contains 22 features, including the test time, sex, and age. Table 1 categorizes these features. The remaining features are the data related to the voice recordings. While abnormalities in any of these features could be symptomatic of many causes, they also provide measurements for several symptoms of Parkinson’s disease.

The data contains four features that measure jitter and six that measure shimmer. Jitter measures the fluctuations in pitch while shimmer indicates fluctuations in amplitude. Common symptoms of Parkinson's disease include difficulty maintaining pitch as well as speaking softly. Therefore, measurements of both jitter and shimmer can be used to detect and measure these symptoms and can be useful to map the voice data to a UPDRS score.
\begin{table}
\caption{Dataset Attributes}
\begin{center}
\begin{tabular}{p{2.5cm} p{9cm}}
\hline
\multicolumn{1}{l}{\rule{0pt}{12pt}
                   Attribute}&\multicolumn{1}{l}{Brief description}\\[2pt]
\hline\rule{0pt}{12pt}Subject & Integer that uniquely identifies each subject (not used in training/test)\\
Age & Subject age\\
Sex & Subject gender '0' - male, '1' - female\\
Test time & Time since recruitment into the trial. The integer part is the number of days since recruitment.\\
Motor UPDRS	& Clinician's motor UPDRS score, linearly interpolated\\
Total UPDRS	& Clinician's total UPDRS score, linearly interpolated (not used)\\
Jitter (\%),\newline 
Jitter(Abs),\newline Jitter:RAP,\newline Jitter:PPQ5,\newline Jitter:DDP & 
Several measures of variation in fundamental frequency\\
Shimmer,\newline 
Shimmer(dB),\newline Shimmer:APQ3,\newline Shimmer:APQ5,\newline Shimmer:APQ11,\newline Shimmer:DDA & Several measures of variation in amplitude\\
NHR,HNR	& Two measures of ratio of noise to total components in the voice\\
RPDE & A nonlinear dynamical complexity measure\\
DFA	& Signal fractal scaling exponent\\
PPE	& A nonlinear measure of fundamental frequency variation\\[2pt]
\hline
\end{tabular}
\end{center}
\end{table}

The recurrence period density entropy (RPDE) measures deviations in the periods of time-delay embedding of the phase space. When a signal recurs to the same point in the phase space at a certain time, it has a recurrence period of that time. Deviations in periodicity can indicate voice disorders, which may occur as a result of Parkinson’s disease [1].

Noise-to-harmonics and harmonics-to-noise ratios are derived from estimates of signal-to-noise ratio from the voice recording. Detrended fluctuation analysis (DFA) measures the stochastic self-similarity of the noise in the speech sample. Most of this noise is from turbulent airflow through the vocal cords [1].  Each of these measures can capture the breathiness in speech that can be a symptom of Parkinson’s disease.

A common symptom of Parkinson’s disease is an impaired ability to maintain pitch during a sustained phonation. While jitter detects these changes in pitch, it also measures the natural variations in pitch that all healthy people exhibit. It can be difficult for jitter measurements to distinguish between these two types of pitch variations. Pitch Period Entropy (PPE) is based on a logarithmic scale rather than a frequency scale, and it disregards smooth variations [1]. Therefore, it is better suited to detect dysphonia-related changes of pitch.

\subsection{Document Organization}

The next section gives an overview regarding the previous research that has been done on this dataset and a similar dataset. Then in the section titled Machine Learning Strategies and Our Research Road-map, you can find the details about feature analysis and selection, a brief description of the methods that we applied, followed by the different approaches and their results. The last section, discussion and conclusion, is about future works and summarization of our study on this dataset.

\section{Related Work}

Approximately 60,000 Americans are diagnosed with Parkinson’s disease each year, but only 4\% of patients are diagnosed before the age of 50 [1] [2] [3]. In order to diagnose Parkinson’s disease earlier, there have been several works, based on a dataset of voice recordings of patients with Parkinson’s disease and healthy patients. First, work by Max Little on this initial dataset predicts whether a subject is healthy or has Parkinson's disease using phonetic analysis of voice recordings to measure dysphonia [1]. 
Much similar work has been done to create models that can accurately predict whether a person has Parkinson's disease or is healthy based on the phonetic analysis of voice recordings. Max Little et al. introduced the dataset and found success with SVM (Support Vector Machine), which indicated that voice measurements can be a suitable way to diagnose Parkinson's disease [1].  Further research has contributed methods to solve this problem with various machine learning techniques [5],which successfully used Artificial Neural Networks as well as a neuro-fuzzy classifier. The neuro-fuzzy classifier achieved a high accuracy on the testing set for this binary classification problem. The work to classify people as healthy or having Parkinson's disease has favorable results, but due to unbalanced data available, we cannot know whether a model is reliable. 
The subsequent work was on a more complex dataset, to monitor patients with early-stage Parkinson's disease [2], which is the basis for this paper.  Age, sex, and voice are all important factors to identify Parkinson's, or in other words, are used by clinicians to calculate a UPDRS score. Therefore, both the above ideas seem promising for identification and monitoring of people with Parkinson's disease. 
As for this paper, we are more concerned with the more recent data and work [2], which attempts to map data from voice recordings to UPDRS scores. The original authors have tackled this problem as a regression problem, by using logistic regression and CART (Classification and Regression Tree). In their work, they have mentioned that by using the CART method, they could reduce the mean absolute error to 4.5 on the training set and 5.8 on the testing set. However, in this study, even with 10-fold cross-validation, we achieved a mean absolute error even lower than 1.9.

\section{Machine Learning Strategies and Our Research Road-Map}

In this section, we examine at first the correlation of the features to the class and discuss feature selection. Next, we describe the machine learning techniques that we applied to this dataset in different manners. We applied regression methods to the data with a continuous class in an attempt to map the phonetic features to the severity of the symptoms from Parkinson's disease. We also undertook this as a classification problem and tried various classification techniques.

\subsection{Feature Selection}

The 22 features accessible in the dataset had already been selected from 132 attributes of the raw data, which is not available to the public [2]. We selected 18 features for this study after excluding the subject, test time, and total UPDRS features from the 22 available ones. We selected the motor UPDRS as the target attribute. Also, motor UPDRS has a high correlation with total UPDRS, so we used motor UPDRS as our sole class. Table 2 shows the correlation between each of those 18 features with motor UPDRS, ordered by correlation.

\begin{table}
\caption{Features correlation with motor UPDRS}
\begin{center}
\begin{tabular}{l l l}
\hline
\multicolumn{1}{l}{\rule{0pt}{12pt}Rank}&\multicolumn{1}{l}{  Correlation  }&\multicolumn{1}{l}{  Attribute}\\[2pt]
\hline\rule{0pt}{12pt}1	&0.2737	&Age\\
2	&0.1624	&PPE\\
3	&0.1366	&Shimmer:APQ11\\
4	&0.1286	&RPDE\\
5	&0.1101	&Shimmer:dB\\
6	&0.1023	&Shimmer\\
7	&0.0921	&Shimmer:APQ5\\
8	&0.0848	&Jitter\\
9	&0.0843	&Shimmer:APQ3\\
10	&0.0843	&Shimmer:DDA\\
11	&0.0763	&Jitter:PPQ5\\
12	&0.075	&NHR\\
13	&0.0727	&Jitter:DDP\\
14	&0.0727	&Jitter:RAP\\
15	&0.0509	&Jitter:Abs\\
16	&0.0312	&Sex\\
17	&-0.1162	&DFA\\
18	&-0.157	&HNR\\[2pt]
\hline
\end{tabular}
\end{center}
\end{table}

\subsection{Regression-Based Methods}

We created regression models to determine a relationship between the features and the continuous class (motor UPDRS). We tried many different techniques in different categories of machine learning methods, including trees, functions, multilayer perceptron, and instance-based learning. Table 3 shows the correlation coefficient and mean absolute error for the top performing regression models. We omitted results that did not compare well to the top models.

Overall, the tree-based regression models performed the best. The previous work on this data measured their results by using the mean absolute error. For this reason, we also observed the mean absolute error for each method, so we could meaningfully compare our results to the previous works. For more insight into the results, we also noted the correlation coefficient.

\begin{table}
\caption{Regression-based methods results}
\begin{center}
\begin{tabular}{l c c}
\hline
\multicolumn{1}{l}{\rule{0pt}{12pt}Method}&\multicolumn{1}{l}{Correlation coefficient}&\multicolumn{1}{l}{   Mean absolute error}\\[2pt]
\hline\rule{0pt}{12pt}M5P tree	&0.9463	&1.9285\\
SVM (nu-SVR)	&0.9335	&2.0612\\
REPTree	&0.9282	&2.0157\\
k-NN	&0.8619	&2.8239\\[2pt]
\hline
\end{tabular}
\end{center}
\end{table}

M5 model tree [6] combines Decision and Regression tree. M5 model (M5P) first constructs a decision tree, and each leaf of that tree is a regression model. Therefore, the output we get out of the M5 model are real values instead of classes. As our dataset’s dependent variable has continuous values, we chose this method. It is different from a Classification and Regression Tree (CART) method as CART generates either a decision or regression tree based on the type of dependent variable and M5 model tree uses both techniques together. This method is one of the best performers.

Support Vector Machines (SVMs) [7] are another popular machine learning algorithm that can handle both classification and regression with the detection of outliers. SVM tries to find an optimal hyperplane that categorizes the new inputs.  We attempted two methods: epsilon-SVR and nu-SVR. There are multiple kernels to find the optimal hyperplane, including linear, radial, sigmoidal, and polynomial. Since the data was complex and has a high dimensionality, radial kernel tends to work faster and better than any other kernel type. We found that nu-SVR performed better than epsilon-SVR for this data. The nu-SVR method got 0.9335 as the highest correlation coefficient, only marginally higher than epsilon-SVR, which achieved 0.9301 for the correlation coefficient. However, the mean absolute error of 2.06 was also slightly higher compared to that of epsilon-SVR, 2.02.

Reduced Error Pruning Tree [8], or REPTree, is Weka's implementation of a fast decision tree learner. The REPTree is sometimes preferred over other trees because it prevents the tree from growing linearly with the sample size when growth will not improve the accuracy [9]. It uses information gain to build a regression tree and prunes it with reduced-error pruning. The REPTree for classification yielded a correlation coefficient of 0.92 and a mean absolute error of 2.02.

\subsection{Instance-Based Learning}

Instance-based learning, such as k-nearest neighbors (k-NN) [10], uses a function that is locally approximated and defers computation to the classification or value assignment. Instead of generalizing the data, new instances are assigned values based directly on the training data. As a regression method, k-NN outputs the average of the values of the instance’s nearest neighbors. We applied the k-NN algorithm to the discretized data and classified instances based on seven nearest neighbors. We used Manhattan distance and weighted the distance with one divided by the distance. This method yielded a correlation coefficient of 0.86 and 2.83 as the mean absolute error.

\subsection{Ensemble Methods}

Using Ensemble Methods, we combined several regression techniques with other methods in an attempt to improve upon the best results we achieved. Some of these methods include bagging, boosting, stacking, voting, and Iterative Absolute Error Regression in conjunction with other regression methods.

Bagging [11], or bootstrap aggregation, uses multiple predictions of a method with high variance. Many subsamples of the data are made with replacement, and a prediction is made with a machine learning method for each subsample. The result from bagging is the average of the result of all of these predictions.

Stacking [12] makes use of several machine learning models. We applied multiple methods to the original dataset. There is a metalayer which uses another model that uses the individual results as its input and creates a prediction. Our best stacking result stacked M5P tree and REPTree, and we used M5PTree as the model for the metalayer.

Voting [13], or simple averaging for this regression problem used multiple machine learning methods. The average of their results was the output for the voting algorithm. Our best model with this ensemble method was once again the M5P Tree and the REPTree.

Random Forest [14] is a tree-based ensemble learning method that can be used for both classification and regression. It operates by constructing a multitude of decision trees at training time and outputting class that is the mode of classes (classification) or mean prediction (regression) of the individual trees. These trees are generally fast and accurate but sometimes suffer from over-fitting.

Tree-based methods did provide the best results and tend to work well in ensemble methods due to their high variance. The best performing model was bagging with the M5P tree as the base method. This yielded a correlation coefficient of 0.95 and a mean absolute error of 1.87. Table 4 features some of the best ensemble methods for regression.

\begin{table}
\caption{Regression-based ensemble methods results}
\begin{center}
\begin{tabular}{p{4cm} c c}
\hline
\multicolumn{1}{l}{\rule{0pt}{12pt}Ensemble method}&\multicolumn{1}{l}{Correlation coefficient}&\multicolumn{1}{l}{   Mean absolute error}\\[2pt]
\hline\rule{0pt}{12pt}Bagging M5P Tree	&0.9529	&1.8674\\
Stacking M5P Tree with REPTree by M5P Tree	&0.9502	&1.8563\\
Vote M5P Tree and REPTree	&0.9465	&1.9163\\
Random Forest	&0.9167	&2.7372\\[2pt]
\hline
\end{tabular}
\end{center}
\end{table}

\subsection{Verification}

In order to confirm that our results are consistent with real values for the motor UPDRS scores and not just fitting to the linear interpolation, we also tried the regression techniques on a subset of the data. In this subset, we used only instances in which the value for the motor UPDRS was a whole number. While some interpolated instances may coincidentally have a whole number for the motor UPDRS, this method ensured that much of the subset was the data where a clinician examined the patient and calculated a UPDRS score.

The results we achieved with this subset of the data surpassed those of the whole dataset. In particular, the M5P model has a correlation coefficient of 0.95 and a mean absolute error of 1.87. These results indicate that our findings from regression methods are reliable and are not simply fitting to the linearly interpolated data.

\subsection{Classification by Discretization}

In addition to the regression models, we also attempted other methods of modeling the data. Below we briefly summarize these different approaches to this problem since the results were not promising.

To classify the instances into meaningful intervals, we discretized the data. According to the work of Pablo Martinez-Martin et al. [15], UPDRS scores can be used to indicate the severity of the disease. These authors classified motor UPDRS scores from 1 to 32 as mild, 33 to 58 as moderate, and 59 and above as severe. Using these intervals, the data we used had 5,254 instances with mild Parkinson’s disease and only 621 instances that were moderate, and no severe cases.
 
We used multiple tree-based methods, including C4.5, Classification and Regression Tree, and Logit Boost Alternating Decision Tree. Weka's J48 algorithm [16] is essentially an implementation of the C4.5 algorithm (a type of decision tree methods). C4.5 is an improvement over the ID3 algorithm. J48 is capable of handling discrete as well as continuous values. It also has an added advantage of allowing a pruned tree. Our data has attributes with continuous values, so we chose the J48 classifier as one of the methods and tried multiple configurations.

SimpleCART [17] is a type of Classification and Regression Tree. The classification tree predicts the class of the dependent variable, while the regression tree outputs a real number.  The SimpleCART technique produces either a classification or regression tree based on whether the dependent variable is categorical or numeric, respectively. The class attribute of our dataset is of numeric type, but after discretizing the class label into bins, we converted it to a categorical type, so the SimpleCART algorithm treated our discretized dataset as a classification problem.

LADTree (Logit boost Alternating Decision tree) [18] is a type of alternating decision tree for multi-class classification.  Alternating Decision Tree was designed for binary classification. ADTrees can be merged into a single tree; therefore, a multiclass model can be derived by merging several binary class trees using some voting model. LADTree uses LogitBoost strategy for boosting. In simple terms, boosting gives relatively more weight to misclassified instances compared to correctly classified instances for the next iteration of boosting. Generally, the boosting iteration is directly proportional to the number of iterations.

Bayes Net [19] is a probabilistic directed acyclic graphical model. This model represents a set of variables and their conditional dependencies through a direct acyclic graph. Each node’s output depends on the particular set of values of its parent nodes. The nodes which are not connected to each other are considered as conditionally independent nodes from each other. Unlike the Naïve Bayes [20] assumption of conditional independence, Bayesian Belief networks describe conditional independence among a subset of variables.

K-nearest neighbors (K-NN) [10], can also be used for both classification and regression problems. When used for classification, k-NN classifies a new instance with the class held by the majority of its nearest neighbors. We applied the K-NN algorithm [10] to the discretized data and classified the motor UPDRS values for instances based on six nearest neighbors and measuring using Manhattan distance.

The Multilayer Perceptron (MLP) is a type of artificial neural network [21] [22]. An MLP consists of one or more layers with a different number of nodes in each, called the network architecture. Using some activation function such as sigmoid in each node combined with the backpropagation technique makes such networks useful for machine learning classification and regression tasks. For this dataset we applied variant network architectures, single-layer to five-layer networks with a range of 1 to 10 nodes in each.

The state of the art, known as deep learning, is the developed MLP into more layers containing more nodes with more options of activation functions and training algorithms [23]. We tried several different architectures of deep neural network (DNNs) using Keras [24] and Tensorflow [25]. The results were not promising in comparison to listed models.

After all, the classification results did not compare well to the regression-based methods. We speculate that this is because of the unbalanced data.

\subsection{Multi-Instance Learning}

We also used multi-instance learning [26]. For each subject in this dataset, there are approximately six voice recordings for each time step. Rather than considering each of these recordings by itself, multi-instance learning collects these instances into bags. Each bag holds one person’s voice recording data that was taken at the same time step, and every instance in the bag is assigned to the same class. We have 995 bags for the 42 subjects, amounting to about 24 per subject, or one per week.

We propositionalized the bags, creating one instance for each bag with the mean of the values of the aggregated instances. This creates 995 instances, one for each person at every time interval. We then were able to apply single-instance classifiers, including Bayesian methods, decision trees, SVMs, and Multi-Layer Perceptron.  These methods yielded similar results to those that we achieved with classification using the data with all 5,875 instances.

The results of this approach were similar to those of the classification models and not significant compared to the regression results.

\section{Discussion and Conclusion}
Our results from the classification problem indicate that these measures of dysphonia may be used to determine the severity of the symptoms of Parkinson's disease. Results with higher accuracy as well as a better ability to predict the minority class suggest a higher likelihood that model can be accurately used with more diverse data. However, we found the regression results to be even more promising. We favored regression over classification with this data because, in order to classify, we must discretize into bins. Meanwhile, regression techniques can map the UPDRS score to a more precise value. It is more meaningful for a model to output a motor UPDRS score of 15 than to say it is in the range of 0 to 16. Our best performing regression method was bagging using the M5P model tree as the base method, which achieved a correlation coefficient of 0.95 and a mean absolute error of 1.86.

To our knowledge, the only existing work on this data belongs to the same authors who created this dataset [2]. We tried many regression methods and compared them to the findings of these authors. We only compiled our significant findings in this report.  Our regression methods resulted in high correlation coefficients. However, the previous work made no mention of the correlation coefficient, so we used the mean absolute error to compare our results. The best results from this earlier work had a mean absolute error of 4.5 on the training set and 5.8 on the testing set. We were able to lower the mean absolute error to 1.9, a significant decrease from the earlier work's mean absolute errors. We attained the lower mean absolute error despite using 10-fold cross-validation, which makes our results more reliable. These models indicate that motor UPDRS can be calculated using these voice measurements in the early stages of the disease with even more precision than previously thought. These results suggest that voice recordings may be a reliable approach to monitoring Parkinson's disease from the comfort of the patient's home, potentially reducing the frequency of doctor visits and giving patients more freedom with their time.

There are many possibilities of future work on this type of data. Researchers are currently collecting more data of this type. If more data is collected from patients at all stages of Parkinson’s, similar techniques could be applied to determine whether vocal parameters can still be mapped to UPDRS scores at a later stage of the disease. Additionally, data from the later stages of Parkinson’s could be used to attempt to predict the progression of the disease.

As a result of our comparative study of machine learning methods, we discovered that the new methods of deep learning are not as efficient and competitive as trees for many tabular data. A data scientist needs to know about all machine learning methods and different types of datasets to achieve the best accuracy and efficiency. The results of our comparative study of variant machine learning models supports the same claim of [27][28][29] that for many tabular data, the older models (e.g., decision trees) outperform cutting edge deep learning models. Also, we should consider that some machine learning models such as decision trees are much faster than DNNs and could be run in very simple machines like on edge devices (e.g., cellphones).

\vspace{3pt}
\textbf{Acknowledgement}

Here we want to appreciate the help of Pawan Yadav and Ankit Joshi for part of the early implementation of the experiments.
%
%

\end{document}